\documentclass[prl,twocolumn,nofootinbib,reprint]{revtex4-1}
\usepackage{amsmath,amssymb,amsthm,bm,bbm}
\usepackage{graphicx}
\usepackage{color} 
\usepackage{hyperref}
\usepackage{cancel}
\usepackage{slashed}

\theoremstyle{definition}

\allowdisplaybreaks[2]

\definecolor{mypink1}{rgb}{0.858, 0.188, 0.478}

\begin{document}

\hfill OU-HET-1139

\vspace{0.3cm}

\title{Muon Electric Dipole Moment as a Probe of Flavor-Diagonal CP Violation
\\
}
\author{Yuichiro Nakai$^{1}$, Ryosuke Sato$^{2}$ and Yoshihiro Shigekami$^{1}$}
\affiliation{\vspace{2mm} \\
$^1$Tsung-Dao Lee Institute and School of Physics and Astronomy, \\Shanghai
Jiao Tong University, 800 Dongchuan Road, Shanghai 200240, China \\
$^2$Department of Physics, Osaka University, 
Toyonaka, Osaka 560-0043, Japan
}

\begin{abstract}
Electric dipole moments (EDMs) of elementary particles are powerful probes of new physics with flavor-diagonal CP violation.
The reported discrepancy in the muon anomalous magnetic moment motivates us to explore
to what extent new physics with flavor-diagonal CP violation to address the discrepancy is probed
by searches for the muon EDM.
As a benchmark, we focus on a CP-violating two-Higgs-doublet model to explain the muon $g-2$ anomaly
where the muon exclusively couples to one Higgs doublet.
Since contributions to flavor violating processes as well as the electron EDM are suppressed,
the muon EDM becomes an essential probe of the model.
Our result shows that some viable parameter space leads to the muon EDM of
around $d_{\mu} \simeq 6 \times 10^{-23} \, e \, \rm cm$
probed by the PSI experiment and most of the parameter space is covered by the proposed J-PARC experiment.
\end{abstract}


\maketitle

{\bf Introduction.--}
Flavor-diagonal CP violation, which does not involve flavor violating interactions,
is one of the most important portals of physics beyond the Standard Model (SM).
It is insensitive to the Cabibbo-Kobayashi-Maskawa (CKM) phase,
the only confirmed CP violation so far,
and its discovery directly indicates the existence of new physics. 
Electric dipole moments (EDMs) of leptons, nucleons, atoms and molecules
provide the most sensitive probes of physics with such flavor-diagonal CP violation.
Searches for the neutron EDM put a stringent upper limit on
the absolute value of the QCD vacuum angle
\cite{Baker:2006ts,Pendlebury:2015lrz}
and inspire many ideas of physics beyond the SM to solve the strong CP problem.
The electron EDM is immune to QCD effects, which enables the precise comparison between theory and experiment,
and its precise measurement \cite{ACME:2018yjb} constrains a wide range of new physics models with flavor-diagonal CP violation
such as supersymmetry (see $e.g.$ refs.~\cite{Nakai:2016atk,Cesarotti:2018huy}).

The experimental data to measure the muon anomalous magnetic moment $a_\mu \equiv (g-2)/2$
shows the discrepancy between theory and experiment,
$\Delta a_\mu^{\rm obs} = a_\mu^{\rm exp} - a_\mu^{\rm theory} = (25.1 \pm 5.9) \times 10^{-10}$
\cite{Muong-2:2006rrc,Keshavarzi:2018mgv,Muong-2:2021ojo}
(see $e.g.$ ref.~\cite{Keshavarzi:2021eqa} for a review),
which may indicate the existence of physics beyond the SM
contributing to the muon $g-2$ at or below the TeV scale.
Then, it is natural to imagine that the same new physics contribution has the imaginary part and generates a muon EDM.
The current upper limit on the muon EDM is given by
$|d_\mu| < 1.8 \times 10^{-19} \, e \, \rm cm$
\cite{Muong-2:2008ebm}.
The EDM measurements of heavy atoms and molecules indirectly give a stronger bound,
$|d_\mu| < 2 \times 10^{-20} \, e \, \rm cm$
\cite{Ema:2021jds}.
They are still orders of magnitude larger than what we naively expect from
the observed value of the muon $g-2$.
However, there are ongoing and future experiments to measure the muon EDM much more precisely.
The Fermilab $g-2$ experiment to search for the muon EDM will reach a sensitivity down to
$10^{-21} \, e \, \rm cm$
\cite{Chislett:2016jau}.
The J-PARC experiment to measure the muon $g-2$/EDM will achieve the similar sensitivity
\cite{Abe:2019thb}.
In addition, the planned dedicated experiment at PSI will achieve the sensitivity of $6 \times 10^{-23} \, e \, \rm cm$
\cite{Adelmann:2021udj,Sakurai:2022tbk,muonEDMinitiative:2022fmk}.
Finally, the proposal of a dedicated experiment at J-PARC claims to search for the muon EDM at 
the level of $10^{-24} \, e \, \rm cm$
\cite{Farley:2003wt}.
These situations motivate us to explore
to what extent new physics to explain the muon $g-2$ anomaly is probed by searches for the muon EDM.

If a CP-violating new physics model follows minimal flavor violation (MFV)
\cite{Chivukula:1987fw,Hall:1990ac,Buras:2000dm,DAmbrosio:2002vsn,He:2014uya},
the current upper limit on the electron EDM
\cite{ACME:2018yjb}
predicts $|d_\mu| = |(m_\mu / m_e) d_e| < 2.3 \times 10^{-27} \, e \, \rm cm$,
which is even far smaller than sensitivities expected at future dedicated searches.
On the other hand, there are two possibilities that can lead to a large muon EDM.
The first possibility is to have flavor-off-diagonal CP violation
\cite{Hiller:2010ib,Omura:2015xcg,Abe:2019bkf,Hou:2021zqq}.
In this case, the muon EDM is not necessarily suppressed by the muon mass $m_{\mu}$.
However, such models generate a variety of flavor violating processes
and the muon EDM cannot be a unique probe of the models (other than direct collider searches).
The other possibility is to have flavor-diagonal CP violation which only the muon can access
so that the electron EDM is significantly suppressed.
Although that interesting possibility has been pointed out in the literature
(see ref.~\cite{Crivellin:2018qmi}),
a simple model without ad hoc assumptions is still lacking
and implications of future muon EDM experiments for that case are unclear.

\begin{table*}[!t]
\centering
\begin{tabular}{|c|c|c|c|c|c|c|c|c|c|c|c|}
\hline
 & $q_L^a$ & $u_R^a$ & $d_R^a$ & $\ell_L^e$ & $\ell_L^{\tau}$ & $\ell_L^{\mu}$ & $e_R$ & $\tau_R$ & $\mu_R$ & $\Phi_1$ & $\Phi_2$ \\ \hline
SU(3)$_C$ & {\bf 3} & {\bf 3} & {\bf 3} & {\bf 1} & {\bf 1} & {\bf 1} & {\bf 1} & {\bf 1} & {\bf 1} & {\bf 1} & {\bf 1} \\
SU(2)$_L$ & {\bf 2} & {\bf 1} & {\bf 1} & {\bf 2} & {\bf 2} & {\bf 2} & {\bf 1} & {\bf 1} & {\bf 1} & {\bf 2} & {\bf 2} \\
U(1)$_Y$ & 1/6 & 2/3 & $-1/3$ & $-1/2$ & $-1/2$ & $-1/2$ & $-1$ & $-1$ & $-1$ & 1/2 & 1/2 \\
$Z_4$ & 1 & 1 & 1 & 1 & 1 & $i$ & 1 & 1 & $i$ & $-1$ & 1 \\ \hline
\end{tabular}
\caption{The charge assignments of fermions and Higgs doublets in the model. The superscript $a(=1,2,3)$ is the generation index.}
\label{tab:Z4}
\end{table*}

In the present letter,
we provide a complete benchmark model to illustrate
to what extent new physics to address the muon $g-2$ anomaly is probed by searches for the muon EDM
without being troubled with other CP and flavor observables.
Our focus is on the two-Higgs-doublet model (2HDM)
\cite{Lee:1973iz} (for a review, see ref.~\cite{Branco:2011iw}),
which is one of the simplest extensions of the SM that may explain the muon $g-2$ anomaly
\cite{Cao:2009as,Broggio:2014mna,Wang:2014sda,Abe:2015oca,Omura:2015xcg,Chun:2015hsa,Abe:2017jqo,Abe:2019bkf}.
We then consider a CP-violating 2HDM
where a softly-broken discrete symmetry makes it possible that the muon exclusively couples to one Higgs doublet.
This model only contains flavor-diagonal CP violation which only the muon can effectively access
and suppresses contributions to flavor violating processes as well as the electron EDM.
The muon EDM hence becomes an essential probe other than direct collider searches.

{\bf CP-violating muon specific 2HDM.--}
We consider two Higgs doublet fields $\Phi_{\alpha} \, (\alpha = 1,2)$ where the muon exclusively couples to one Higgs doublet $\Phi_1$
\cite{Abe:2017jqo}. 
This is justified by imposing a softly-broken $Z_4$ symmetry
under which SM fermions and Higgs doublets transform as shown in Tab.~\ref{tab:Z4}. 
Then, Yukawa interactions are given by
\begin{align}
\mathcal{L}_Y &= - \bar{q}_L \widetilde{\Phi}_2 Y_u u_R - \bar{q} \Phi_2 Y_d d_R \nonumber \\[0.5ex]
&\hspace{1.2em}- \bar{L}_L \Phi_1 Y_{\ell 1} E_R - \bar{L}_L \Phi_2 Y_{\ell 2} E_R + {\rm h.c.} \, ,
\label{eq:Yukawa}
\end{align}
where $Y_u, Y_d, Y_{\ell 1}, Y_{\ell 2}$ are $3 \times 3$ Yukawa matrices,
$\widetilde{\Phi}_2 \equiv i \sigma^2 \Phi_2^*$ and $L_L, E_R$ in our notation are
\begin{align}
L_L = \left( \ell_L^e, \ell_L^{\tau}, \ell_L^{\mu} \right)^T  , \quad E_R = \left( e_R, \tau_R, \mu_R \right)^T  ,
\end{align}
and hence $Y_{\ell 1}, Y_{\ell 2}$ are written as
\begin{align}
Y_{\ell 1} = {\rm diag} ( 0, 0, y_{\mu} ) , \quad Y_{\ell 2} = {\rm diag} ( y_e, y_{\tau}, 0 ) \, .
\end{align}
Although off-diagonal elements in $Y_{\ell 2}$ are not forbidden by symmetries,
they are taken to be zero by field rotations without loss of generality~\footnote{Neutrino mixings
between $\nu_{\mu}$ and $\nu_{e, \tau}$ are forbidden by the $Z_4$ symmetry.
However, we can introduce a triplet scalar with hypercharge $-1$ and $Z_4$ charge $-i$ so that
these mixings are generated. The detailed discussion is given in ref.~\cite{Abe:2017jqo}.
This triplet scalar does not affect our result if it is heavier than the other particles in the model. 
}. 
The Higgs potential is given by
\begin{align}
V_{\Phi} &= m_{11}^2 \Phi_1^{\dagger} \Phi_1 + m_{22}^2 \Phi_2^{\dagger} \Phi_2 - \bigl[ m_{12}^2 \Phi_1^{\dagger} \Phi_2 + {\rm h.c.} \bigr] \nonumber \\[0.5ex]
&\hspace{1.2em} + \frac{\lambda_1}{2} \bigl( \Phi_1^{\dagger} \Phi_1 \bigr)^2 + \frac{\lambda_2}{2} \bigl( \Phi_2^{\dagger} \Phi_2 \bigr)^2 \nonumber \\[0.5ex]
&\hspace{1.2em} + \lambda_3 \bigl( \Phi_1^{\dagger} \Phi_1 \bigr) \bigl( \Phi_2^{\dagger} \Phi_2 \bigr) + \lambda_4 \bigl( \Phi_1^{\dagger} \Phi_2 \bigr) \bigl( \Phi_2^{\dagger} \Phi_1 \bigr) \nonumber \\[0.5ex]
&\hspace{1.2em} + \left[ \frac{\lambda_5}{2} \bigl( \Phi_1^{\dagger} \Phi_2 \bigr)^2 + {\rm h.c.} \right] \, .
\label{eq:Vori}
\end{align}
Here, $m_{11}^2, m_{22}^2$ and $\lambda_{1,2,3,4}$ are real parameters
while $m_{12}^2$ and $\lambda_5$ can be complex.
Only a nonzero $m_{12}^2$ breaks the $Z_4$ symmetry softly.
The Higgs fields are parameterized as
\begin{align}
\Phi_{\alpha} = \begin{pmatrix}
\pi_{\alpha}^+ \\
\frac{1}{\sqrt{2}} \left( v_{\alpha} + h_{\alpha} + i a_{\alpha} \right)
\end{pmatrix} \, ,
\label{eq:doubletdef}
\end{align}
where $v_1$ and $v_2$ are vacuum expectation values (VEVs) of $\Phi_1$ and $\Phi_2$, respectively,
and satisfy $v^2 = v_1^2 + v_2^2$ with $v \simeq 246 \, \rm GeV$.
We define the ratio of the VEVs as $\tan \beta \equiv v_2 / v_1$.

A minimization condition of the Higgs potential~\eqref{eq:Vori}
relates the imaginary parts of $m_{12}^2$ and $\lambda_5$ with each other,\footnote{
The analysis of the same Higgs potential has been presented in $e.g.$ refs.~\cite{ElKaffas:2006gdt,Arhrib:2010ju,Cheung:2020ugr,Altmannshofer:2020shb}.
See also refs.~\cite{Davidson:2005cw,Haber:2006ue,Osland:2008aw,Kanemura:2015ska,Boto:2020wyf}
for a more generic Higgs potential.}
\begin{align}
v^2 s_{\beta} c_{\beta} {\rm Im} \lambda_5 = 2 {\rm Im} m_{12}^2 \, ,
\end{align}
where we define $s_{\beta} \equiv \sin \beta$ and $c_{\beta} \equiv \cos \beta$. 
The same manner is applied to other mixing angles. 
Therefore, only one complex phase is a physical CP violation parameter. 
The other minimization conditions reduce the parameters of the Higgs potential,
\begin{align}
m_{11}^2 &= - \frac{1}{2} \lambda_1 v^2 c_{\beta}^2 - \frac{1}{2} \lambda_{345} v^2 s_{\beta}^2 + {\rm Re} m_{12}^2 t_{\beta} \, , \\[1ex]
m_{22}^2 &= - \frac{1}{2} \lambda_2 v^2 s_{\beta}^2 - \frac{1}{2} \lambda_{345} v^2 c_{\beta}^2 + {\rm Re} m_{12}^2 t_{\beta}^{-1} \, ,
\end{align}
where $\lambda_{345} \equiv \lambda_3 + \lambda_4 + {\rm Re} \lambda_5$ is defined.

Let us now discuss the Higgs mass spectrum.
The mass matrix of the neutral Higgs fields $(h_1,h_2,a_1,a_2)$ is written as
$({\cal M}^2)_{xy} = \partial^2 V_{\Phi} / \partial x \partial y$ with $x, y = h_1,h_2,a_1,a_2$.
This mass matrix can be block-diagonalized,
\begin{align}
\widetilde {\cal M}^2 \equiv R_1^T {\cal M}^2 R_1 = \begin{pmatrix}
\delta_{\tilde h \tilde h} & \delta_{\tilde h \tilde H} & \delta_{\tilde h \tilde A} & 0 \\
\delta_{\tilde h \tilde H} & M^2 + \delta_{\tilde H \tilde H} & \delta_{\tilde H \tilde A} & 0 \\
\delta_{\tilde h \tilde A} & \delta_{\tilde H \tilde A} & M^2 + \delta_{\tilde A \tilde A} & 0 \\
0 & 0 & 0 & 0
\end{pmatrix} \, ,
\end{align}
by the rotation matrix, 
\begin{align}
R_1 = \begin{pmatrix}
c_{\beta} & - s_{\beta} & 0 & 0 \\
s_{\beta} & c_{\beta} & 0 & 0 \\
0 & 0 & - s_{\beta} & c_{\beta} \\
0 & 0 & c_{\beta} & s_{\beta}
\end{pmatrix} \, .
\end{align}
Here, we have defined
\begin{align}
&M^2 \equiv \frac{1}{s_{\beta} c_{\beta}} {\rm Re} m_{12}^2 \, , \\[1ex]
&\delta_{\tilde h \tilde h} \equiv \lambda_1 v^2 c_{\beta}^4 + \lambda_2 v^2 s_{\beta}^4 + 2 \lambda_{345} v^2 s_{\beta}^2 c_{\beta}^2 \, , \\[1ex]
&\delta_{\tilde h \tilde H} \equiv -(\lambda_1 - \lambda_{345} ) v^2 s_{\beta} c_{\beta}^3 + (\lambda_2 - \lambda_{345}) v^2 s_{\beta}^3 c_\beta \, , \label{eq:delhH} \\[1ex]
&\delta_{\tilde h \tilde A} \equiv - {\rm Im} \lambda_5 v^2 s_{\beta} c_{\beta} \, , \\[1ex]
&\delta_{\tilde H \tilde H} \equiv (\lambda_1 + \lambda_2 - 2\lambda_{345}) v^2 s_{\beta}^2 c_{\beta}^2 \, , \\[1ex]
&\delta_{\tilde H \tilde A} \equiv \frac{1}{2} (-c_{\beta}^2 + s_{\beta}^2) {\rm Im} \lambda_5 v^2 \, , \\[1ex]
&\delta_{\tilde A \tilde A} \equiv - {\rm Re} \lambda_5 v^2 \, . 
\end{align}
For $M^2 > \delta \sim \lambda v^2$,
we can diagonalize $\widetilde {\cal M}^2$ by the rotation matrix,
\begin{align}
R^T \widetilde {\cal M}^2 R = {\rm diag} (m_h^2, m_{H_1}^2, m_{H_2}^2, 0) \, , \quad
R \equiv R_2 R_3  \label{eq:diagM2neutral} \, ,
\end{align}
where
\begin{align}
R_2 &\simeq \begin{pmatrix}
1           & \delta_{\tilde h \tilde H}/M^2 & \delta_{\tilde h \tilde A}/M^2 & 0 \\
-\delta_{\tilde h \tilde H}/M^2 & 1          & 0          & 0 \\
-\delta_{\tilde h \tilde A}/M^2 & 0          & 1          & 0 \\
0           & 0          & 0          & 1
\end{pmatrix} \, , \label{eq:R2} \\[1ex]
R_3 &= \begin{pmatrix}
1       & 0         & 0        & 0 \\
0       & c_{\theta}  & s_{\theta} & 0 \\
0       & - s_{\theta} & c_{\theta} & 0 \\
0       & 0         & 0        & 1
\end{pmatrix} \, . \label{eq:R3}
\end{align}
The masses of the physical neutral Higgs bosons, $i.e.$
$R_1 R (h, H_1, H_2, G)^T \equiv (h_1, h_2, a_1, a_2)^T$
where $G$ denotes a Nambu-Goldstone (NG) mode,
are approximately given in terms of $\delta$ and $M^2$ as
\begin{align}
m_h^2 &= \delta_{\tilde h \tilde h} + \mathcal{O}(1/M^2) \, , \label{eq:mhdef} \\[1ex]
m_{H_1}^2 &\equiv m_H^2 \nonumber \\
&=M^2 + \delta_{\tilde H \tilde H} c_{\theta}^2 - 2 \delta_{\tilde H \tilde A} s_{\theta} c_{\theta} + \delta_{\tilde A \tilde A} s_{\theta}^2 + \mathcal{O}(1/M^2) \, , \label{eq:mH1def} \\
m_{H_2}^2 &\equiv m_H^2 + \Delta m_H^2 \nonumber \\
&= M^2 + \delta_{\tilde H \tilde H} s_{\theta}^2 + 2 \delta_{\tilde H \tilde A} s_{\theta} c_{\theta} + \delta_{\tilde A \tilde A} c_{\theta}^2 + \mathcal{O}(1/M^2) \, , \label{eq:mH2def}
\end{align}
and the mixing angle of $R_3$ is
\begin{align}
\tan 2 \theta = \frac{2 \delta_{\tilde H \tilde A}}{\delta_{\tilde A \tilde A} - \delta_{\tilde H \tilde H}} \, .
\end{align}
We identify the lightest mode $h$ as the observed Higgs boson, $m_h \simeq 125 \, \rm GeV$.
For the charged Higgs fields,
there exist NG modes ($G^{\pm}$) eaten by the longitudinal components of $W^{\pm}$ bosons,
and the remaining modes ($H^{\pm}$) are physical.
They are related to the original fields in Eq.~\eqref{eq:doubletdef} as
\begin{align}
\begin{pmatrix}
- s_{\beta} & c_{\beta} \\
c_{\beta} & s_{\beta}
\end{pmatrix} \begin{pmatrix}
H^{\pm} \\
G^{\pm}
\end{pmatrix} = \begin{pmatrix}
\pi_1^{\pm} \\
\pi_2^{\pm}
\end{pmatrix} \, .
\label{eq:chargedmode}
\end{align}
The physical charged Higgs mass is given by
$m_{H^{\pm}}^2 = M^2 - \frac{v^2}{2} (\lambda_4 + {\rm Re} \lambda_5)$.

The Yukawa interactions in Eq.~\eqref{eq:Yukawa} are expressed in terms of the physical Higgs bosons as
\begin{align}
\mathcal{L}_Y^{\rm int} &= - \sum_{f \neq \mu} \frac{m_f}{v} \left[ \left( R_{1i} + \frac{R_{2i}}{t_{\beta}} \right) \bar{f} f + i s_f \frac{R_{3i}}{t_{\beta}} \bar{f} \gamma_5 f \right] \phi_i \nonumber \\
&\hspace{1.2em} - \frac{m_{\mu}}{v} \Bigl[ (R_{1i} - R_{2i} t_{\beta}) \bar{\mu} \mu - i R_{3i} t_{\beta} \bar{\mu} \gamma_5 \mu \Bigr] \phi_i \nonumber\\
&\hspace{1.2em}+ \biggl\{ - \frac{\sqrt{2}}{v t_{\beta}} \sum_{a = 1}^3 \bar{u}^a \left( m_{d^a} P_R - m_{u^a} P_L \right) d^a H^+ \biggr. \nonumber \\
&\hspace{1.2em}+ \left[ \frac{\sqrt{2}}{v} t_{\beta} m_{\mu} \bar{\nu}_{\mu} P_R \mu - \frac{\sqrt{2}}{v t_{\beta}} \sum_{\ell \neq \mu} m_{\ell} \bar{\nu}_{\ell} P_R \ell \right] H^+ \nonumber \\
&\hspace{4.8em}\biggl. + {\rm h.c.} \biggr\} \, ,
\label{eq:LagYuk}
\end{align}
where $\phi_i = (h, H_1, H_2)$, $R_{ij}$ is the element of $R$ defined in Eq.~\eqref{eq:diagM2neutral} and $s_f = +1 \, (-1)$ for down-type quarks and charged leptons (up-type quarks).
Note that the neutral scalar couplings have CP-violating contributions
and only the muon has a CP-violating coupling enhanced by a large $\tan \beta$. 
It has been discussed in ref.~\cite{Abe:2017jqo} that $y_{\mu}$ is also enhanced for a large $\tan \beta$
and $y_{\mu} \lesssim 3$ puts an upper bound $t_{\beta} \lesssim 5000$.

Parameterizing $M^2$ and $m_{H^{\pm}}^2$ as
\begin{align}
M^2 &\equiv m_H^2 + s_{\theta}^2 \Delta m_H^2 - 2 \frac{\delta_{\tilde h \tilde H}}{t_{\beta}} - v^2 \frac{X}{t_{\beta}^2} \, , \label{eq:M2def} \\
m_{H^{\pm}}^2 &\equiv m_H^2 + \Delta m_{\pm}^2 \, ,
\end{align}
with an arbitrary number $X$, 
the Higgs quartic couplings $\lambda_i$ can be written as 
\begin{align}
\lambda_1 v^2 &\simeq m_h^2 + X v^2 \, , \label{eq:lam1} \\[1ex]
\lambda_2 v^2 &\simeq m_h^2 \, , \label{eq:lam2} \\[1ex]
\lambda_3 v^2 &\simeq m_h^2 - 2 s_{\theta}^2 \Delta m_H^2 + 2 \Delta m_{\pm}^2 - \delta_{\tilde h \tilde H} t_{\beta} \, , \label{eq:lam3} \\[1ex] \lambda_4 v^2 &\simeq \Delta m_H^2 - 2 \Delta m_{\pm}^2 \, , \label{eq:lam4} \\[1ex]
{\rm Re} \lambda_5 v^2 &\simeq - c_{2 \theta} \Delta m_H^2 \, , \label{eq:Relam5} \\[1ex]
{\rm Im} \lambda_5 v^2 &\simeq s_{2 \theta} \Delta m_H^2 \, . \label{eq:Imlam5}
\end{align}
Here, we have taken the large $t_{\beta}$ limit
and omitted terms of $\mathcal{O}(v^4 / M^2)$ and $\mathcal{O}(v^2 / t_{\beta})$. 
Then, it can be seen that 
\begin{align}
m_H^2, \, \Delta m_H^2, \, \Delta m_{\pm}^2, \, t_{\beta}, \, \theta, \, X, \, \delta_{\tilde h \tilde H} 
\label{eq:indepparas}
\end{align}
are independent parameters to determine the Higgs potential.
The quartic couplings $\lambda_i$ should satisfy the vacuum stability conditions~\cite{Deshpande:1977rw,Klimenko:1984qx,Sher:1988mj,Nie:1998yn,Kanemura:1999xf,Arhrib:2010ju} and the perturbative unitarity conditions~\cite{Kanemura:1993hm,Akeroyd:2000wc,Ginzburg:2005dt,Arhrib:2010ju,Kanemura:2015ska}
as summarized in supplemental materials. 
Roughly speaking, these conditions require $\lambda_i$ to be $\mathcal{O}(1)$, which are realized
when we take $\Delta m_H^2, \Delta m_{\pm}^2 = \mathcal{O}(v^2)$, $X = \mathcal{O}(1)$ and $\delta_{\tilde h \tilde H} = \mathcal{O}(v^2 / t_{\beta})$. 
In addition, a Landau pole should not show up near the weak scale.
To see this, we look at 1-loop renormalization group equations (RGEs).
For a large $\tan \beta$, we need to take account of contributions from $y_{\mu}$.
The $\beta$-functions can be found in refs.~\cite{Abe:2017jqo,Herren:2017uxn,Bednyakov:2018cmx}.
Those theoretical conditions set a cutoff scale $\Lambda_{\rm cutoff}$ for the theory.
In our numerical analyses, we optimize the parameters to maximize $\Lambda_{\rm cutoff}$,
whose procedure is described in supplemental materials.

\begin{figure*}
\begin{minipage}[!t]{\hsize}
\includegraphics[width=0.333\textwidth,bb= 0 0 450 471]{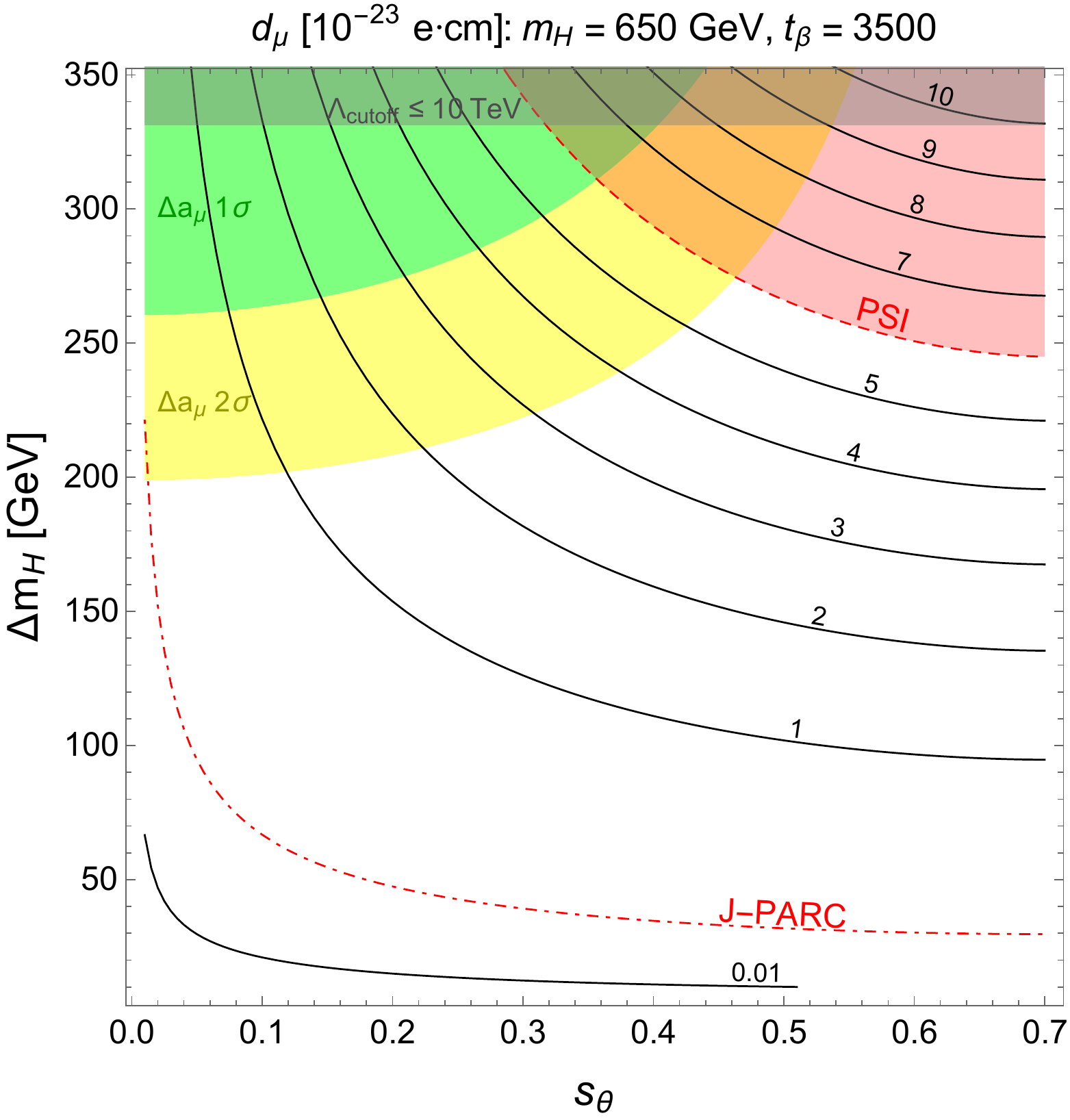}
\hspace{2cm}
\includegraphics[width=0.345\textwidth,bb= 0 0 450 456]{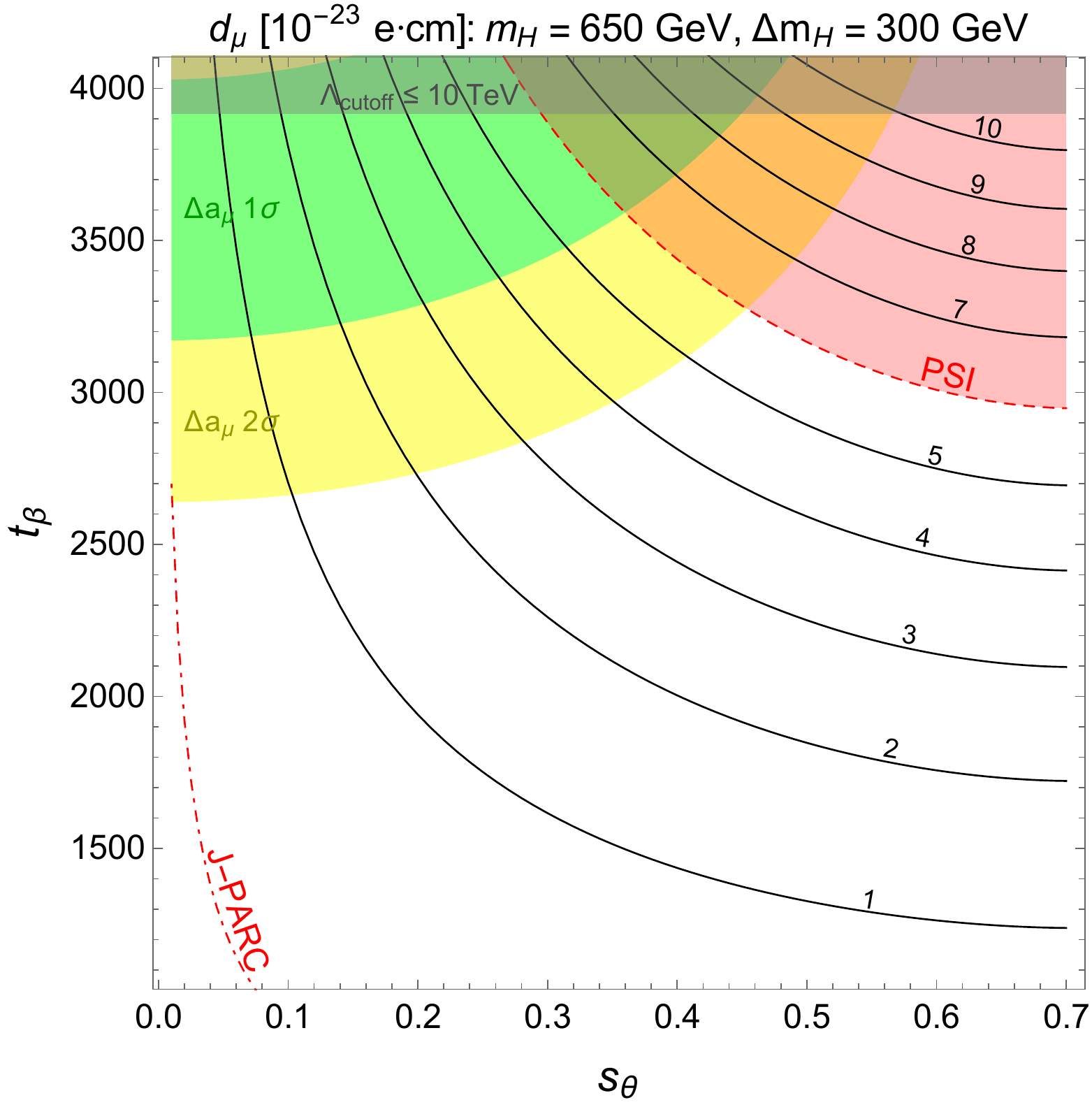}
  \end{minipage}
  \caption{
Contour plots of $d_{\mu}$ and $\Delta a_{\mu}$ for the case of $m_H = 650$ GeV.
{\em Left panel:}
$(s_{\theta}, \Delta m_H)$ plane with $t_{\beta} = 3500$. 
{\em Right panel:}
$(s_{\theta}, t_{\beta})$ plane with $\Delta m_H = 300$ GeV.
In both panels, we take $\Delta m_{\pm}^2 = \Delta m_H^2 / 2$ and define $\Delta m_H \equiv \sqrt{m_{H_2}^2 - m_{H_1}^2}$.
The numbers in the plot correspond to our predictions for $d_{\mu}$ in $10^{-23} \, e$ cm unit. 
The green and yellow bands show our $(g-2)_{\mu}$ prediction with $1\sigma$ and $2\sigma$, respectively. 
The red shaded region denotes the future sensitivity of PSI~\cite{Adelmann:2021udj,Sakurai:2022tbk,muonEDMinitiative:2022fmk},
and the red dash-dotted line is that of J-PARC~\cite{Farley:2003wt}. 
The gray shaded region is excluded by a cutoff scale $\Lambda_{\rm cutoff}$ below 10 TeV. 
}
\label{fig:650GeV} 
\end{figure*}

{\bf Muon $g-2$ and EDM.--}
The muon anomalous magnetic moment receives  
contributions from all the Higgs bosons including the charged Higgs at the one-loop level.\footnote{
As mentioned in ref.~\cite{Abe:2017jqo}, two-loop Barr-Zee type contributions from the 3rd generation fermions are suppressed
because they couple to each scalar with $\cot \beta$. 
Those from a muon loop will be important because the muon couples to each scalar with $\tan \beta$. 
However, these contributions are about two orders of magnitude smaller than one-loop ones
due to the suppression factor $\alpha / \pi$ with loop functions larger than $I_{S,P}(r)$ by $\mathcal{O}(10)$.
Then we ignore these contributions in our analysis.} They are estimated as
\cite{Leveille:1977rc,Haber:1978jt,Krawczyk:1996sm,Dedes:2001nx}
\begin{align}
\Delta a_{\mu}^{\phi_i} &= \frac{m_{\mu}^2}{8 \pi^2 v^2} r_i \left[ (R_{1i} - R_{2i} t_{\beta})^2 I_S(r_i) + R_{3i}^2 t_{\beta}^2 I_P(r_i) \right] \nonumber \\
&\simeq \frac{m_{\mu}^2}{8 \pi^2 v^2} r_i t_{\beta}^2 \left[ R_{2i}^2 I_S(r_i) + R_{3i}^2 I_P(r_i) \right] \, , \label{eq:g-2neutral} \\[1.0ex]
\Delta a_{\mu}^{H^{\pm}} &= \frac{m_{\mu}^2}{8 \pi^2 v^2} t_{\beta}^2 r_{\pm} I_C(r_{\pm}) \, , \label{eq:g-2charged}
\end{align}
where $r_i \equiv m_{\mu}^2 / m_{\phi_i}^2$, $r_{\pm} \equiv m_{\mu}^2 / m_{H^{\pm}}^2$
and $I_{S, P, C}(r)$ are loop functions defined in supplemental materials. 
Note that $I_{S, P}(r)$ are enhanced by two orders of magnitude compared to $I_C(r)$.\footnote{
The part of the reason of this enhancement is that $I_{S,P}$ has a $\log r$ factor and $I_C$ does not have \cite{Dedes:2001nx}.
This logarithmic factor can be understood as the mixing between the dipole operator and
the tree-level four Fermi operator in the effective theory. See, \textit{e.g.}, refs.~\cite{Toharia:2005gm, Gambino:2005eh, Sato:2012xf, Buttazzo:2020ibd}.}
In addition, the SM Higgs contribution $\Delta a_{\mu}^h$ is highly suppressed due to smallness of $R_{2i, 3i}$
(see Eqs.~\eqref{eq:R2}, \eqref{eq:R3}) and no $\tan \beta$-enhancement. 
Then, the dominant contributions to $\Delta a_{\mu}$ are provided by the heavy neutral Higgs bosons $H_{1,2}$. 
Since $I_S(r) \simeq - I_P(r) \sim \log \left( m_H^2 / m_{\mu}^2 \right)$ for $r \ll 1$, $R_{2i} = c_{\theta} \, (- s_{\theta})$ and $R_{3i} = s_{\theta} \, (c_{\theta})$ for $H_1 \, (H_2$),
our prediction of $\Delta a_{\mu}$ is approximately given by
\begin{align}
\Delta a_{\mu} \simeq \frac{m_{\mu}^4}{8 \pi^2 v^2} \frac{\Delta m_H^2}{m_H^4} t_{\beta}^2 c_{2\theta} \log \left( \frac{m_H^2}{m_{\mu}^2} \right) \, .
\label{eq:amurough}
\end{align}
We can see that a sufficiently large $\Delta m_H^2$ and $s_{\theta} \sim 0$ or $\pi$ are required
to explain the muon $g-2$ anomaly.

Since only the neutral scalar couplings contain CP violation,
the muon EDM $d_{\mu}$ receives contributions from only the neutral Higgs bosons at the one-loop level. 
They are estimated as
\begin{align}
d_{\mu}^i &= - \frac{e \, m_{\mu}}{(4 \pi)^2 v^2} r_i \left( R_{1i} - R_{2 i} t_{\beta} \right) R_{3 i} t_{\beta} f_0 (r_i) \nonumber \\[1ex]
&\simeq \frac{e \, m_{\mu}}{(4 \pi)^2 v^2} r_i R_{2 i} R_{3 i} t_{\beta}^2 f_0 (r_i) \, ,
\label{eq:muEDM1loop}
\end{align}
where $f_0 (r)$ is a loop function defined in supplemental materials.
Similar to $\Delta a_{\mu}$,
the contribution from the lightest Higgs $h$ is suppressed 
and then the muon EDM is dominated by the heavy scalar contributions. 
Moreover, using the fact that 
$f_0 (r_2) \simeq f_0 (r_3) \sim \log( m_H^2 / m_{\mu}^2 )$, the sum of one-loop contributions can be expressed by
\begin{align}
d_{\mu} \simeq - \frac{e \, m_{\mu}^3}{32 \pi^2 v^2} \frac{\Delta m_H^2}{m_H^4} t_{\beta}^2 s_{2\theta} \log \left( \frac{m_H^2}{m_{\mu}^2} \right)\, .
\label{eq:dmurough}
\end{align}
Therefore, a small $m_H^2$ and a large $\Delta m_H^2$ are preferred to obtain a large $d_{\mu}$. 
Moreover, $\theta \sim \pi / 4$ is essential to enhance $d_{\mu}$,
which means the maximal mixing between the CP-even and CP-odd scalars.

In addition to the one-loop contributions, two-loop Barr-Zee type diagrams contribute to the muon EDM. 
The expressions of such contributions can be found in refs.~\cite{Abe:2013qla,Nakai:2016atk,Chun:2019oix}. 
In the present case, the diagram with inner fermion loop being the muon 
and outer loop being the photon and a neutral scalar $\phi_i$ gives the dominant contribution.
It is estimated as
\begin{align}
d_{\mu}^{i, \phi_i \mathchar`- \gamma \mathchar`- \gamma} &= - \frac{e \, m_{\mu}}{(4 \pi)^2 v^2} r_i \left( R_{1i} - R_{2 i} t_{\beta} \right) R_{3 i} t_{\beta} \frac{2 \alpha}{\pi} I_{\mu}(r_i) \, ,
\label{eq:muEDMBZ}
\end{align}
where $\alpha \equiv e^2 / 4\pi$ denotes the fine structure constant
and $I_{\mu}(r)$ is a loop function defined in supplemental materials.
Since this loop function is $I_{\mu}(r_i) \sim - (15 \mathchar`- 19) \times f_0(r_i)$
when $m_h \leq m_{\phi_i} \leq 1000$ GeV and $2 \alpha / \pi \simeq 4.6 \times 10^{-3}$, 
the two-loop contribution is one order of magnitude smaller than the one-loop contribution
in Eq.~\eqref{eq:muEDM1loop} and has the opposite sign.

Comparing the expressions of the anomalous magnetic moment in Eq.~\eqref{eq:amurough} and
the EDM in Eq.~\eqref{eq:dmurough}, they have different dependence on $\theta$:
$\Delta a_{\mu} \propto \cos 2 \theta$, while $d_{\mu} \propto \sin 2 \theta$. 
Therefore, $\theta \sim \pi / 8$ is favored to obtain sizable $\Delta a_{\mu}$ and $d_{\mu}$ at the same time.

{\bf Results.--}
Fig.~\ref{fig:650GeV} shows $\Delta a_{\mu}$ and $d_{\mu}$
in $(s_{\theta}, \Delta m_H)$ (left) and $(s_{\theta}, t_{\beta})$ (right) planes for the case of $m_H = 650$ GeV.
It has been shown in ref~\cite{Abe:2017jqo} that searches for extra Higgs bosons at the Large Hadron Collider (LHC)
give a lower bound $m_H \gtrsim 640$ GeV
for the muon specific 2HDM.
We can see that both $\Delta a_{\mu}$ and $d_{\mu}$ are proportional to $\Delta m_H^2 t_{\beta}^2$. 
However, since the dominant contributions to $\Delta a_{\mu}$ and $d_{\mu}$ have the opposite correlation
in terms of $s_{\theta}$ ($\Delta a_{\mu} \propto \cos 2 \theta$ while $d_{\mu} \propto \sin 2 \theta$), 
$s_{\theta} \sim 0$ is favored for $\Delta a_{\mu}$ and $s_{\theta} \sim 0.7$ for $d_{\mu}$. 
We still find an interesting parameter region in each panel where $\Delta a_{\mu}$ is within $1\sigma$ deviation
and $d_{\mu}$ is within the sensitivity of the PSI experiment at the same time.
The other region that can explain the muon $g-2$ anomaly predicts $d_{\mu} = \mathcal{O}(10^{-23}) \, e \, \rm cm$
explored by the J-PARC experiment.

\begin{figure*}
\begin{minipage}[!t]{\hsize}
\includegraphics[width=0.333\textwidth,bb= 0 0 450 471]{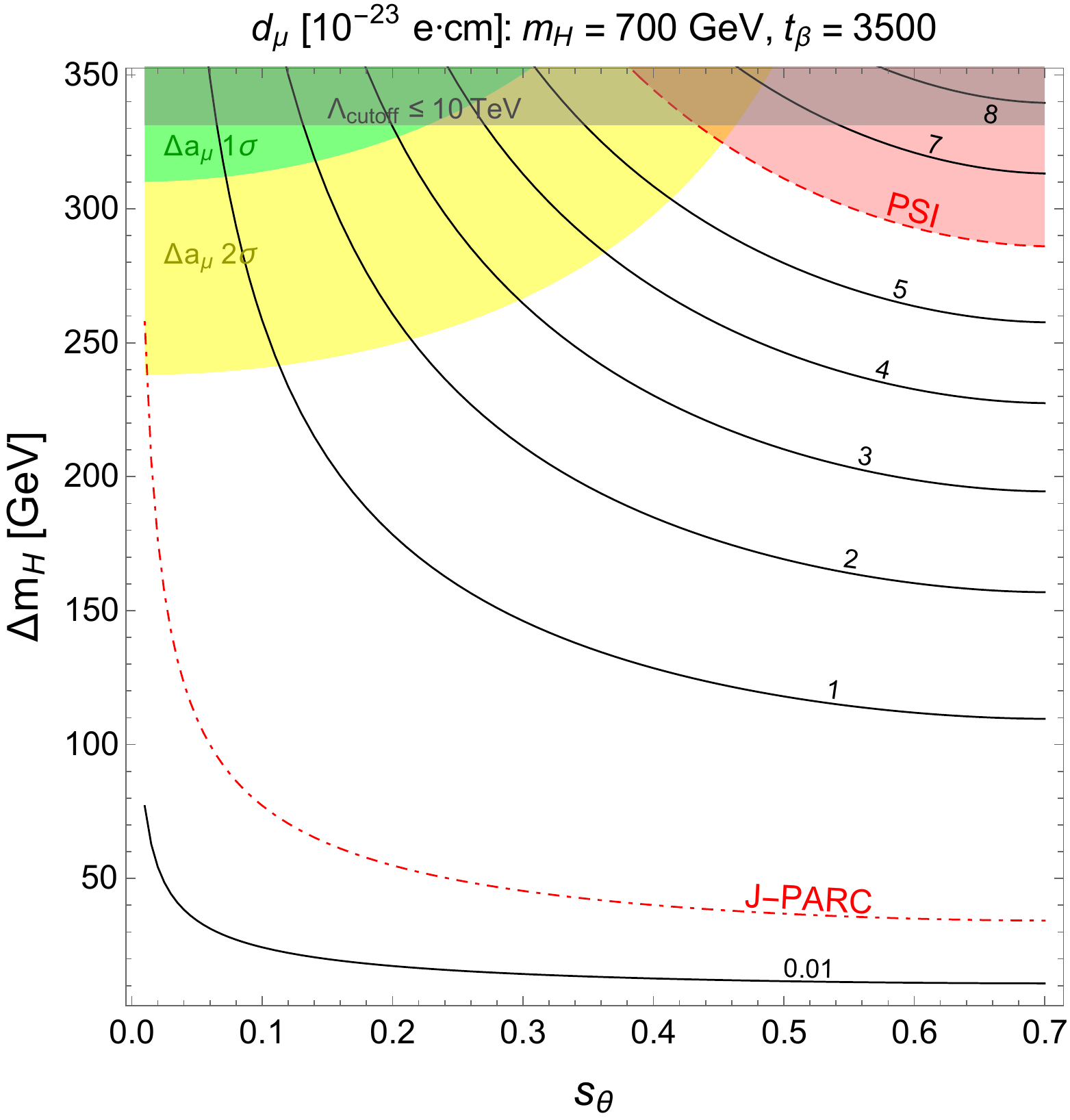}
\hspace{2cm}
\includegraphics[width=0.345\textwidth,bb= 0 0 450 456]{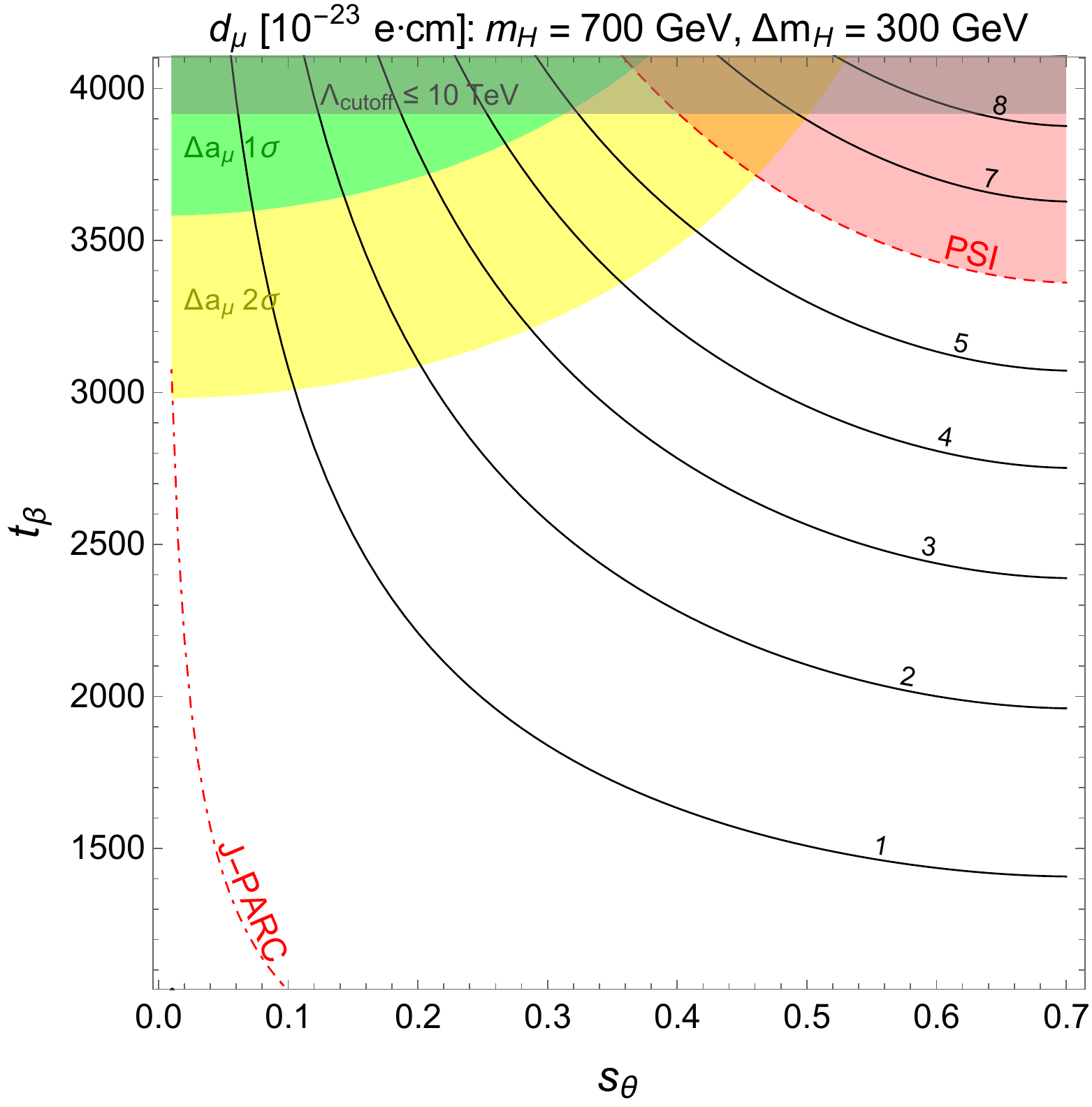}
  \end{minipage}
  \caption{
Contour plots of $d_{\mu}$ and $\Delta a_{\mu}$ for the case of $m_H = 700$ GeV.
The plot manner is the same as that of Fig.~\ref{fig:650GeV}.
}
\label{fig:700GeV} 
\end{figure*}

Fig.~\ref{fig:700GeV} shows the case of $m_H = 700$ GeV.
Since $\Delta a_{\mu}$ and $d_{\mu}$ are proportional to $1/m_H^4$,
they are smaller than those of $m_H = 650$ GeV by a factor of $(650 / 700)^4 \simeq 0.74$. 
As a result, in this case, there is no region which explains the muon $g-2$ anomaly within $1\sigma$
and also gives $d_{\mu}$ within the PSI sensitivity. 
There still exists such a region if we accept a $2\sigma$ deviation for $\Delta a_{\mu}$.
Figs.~\ref{fig:650GeV} and \ref{fig:700GeV} imply that our viable parameter space which can explain the muon $g-2$ anomaly
may be explored first by direct searches for extra Higgs bosons at the LHC
before the PSI experiment finds some signal or puts a bound on the parameter space. 
In other words, if the LHC discovers heavy scalars with mass of $650$-$700$ GeV,
it gives a strong motivation to search for the muon EDM at the PSI experiment to test our model.

The parameter space of interest shown in Figs.~\ref{fig:650GeV} and \ref{fig:700GeV}
is consistent with the current measurements on the $T$ parameter~\cite{ParticleDataGroup:2020ssz}
and the SM-like Higgs coupling to the muon~\cite{ATLAS:2020fzp,CMS:2020xwi}.
The details are summarized in supplemental materials.

A lattice collaboration has claimed that the SM value of $a_\mu^{\rm theory}$ is smaller than
$a_\mu^{\rm exp}$ by only $1.6 \sigma$
\cite{Borsanyi:2020mff}.
If this is the case, choosing $s_{\theta} \sim 0.7$ suppresses $\Delta a_{\mu}$ while it enhances $d_{\mu}$
leading to a signal observed at the PSI experiment.
A heavier $m_H$ can be also consistent with the $g-2$ data.
In this case, taking $\Delta m_H = 300$ GeV, $\Delta m_{\pm}^2 = \Delta m_H^2 / 2$,
$t_{\beta} = 3910$ and $s_{\theta} = 0.7$,
the mass reach of the PSI experiment is $m_H \simeq 760$ GeV
and that of the J-PARC experiment is $m_H \simeq 2210$ GeV. 
In any case, the J-PARC experiment can probe a parameter space beyond the region which can explain
the muon $g-2$ anomaly.

{\bf Conclusions.--}
We have provided a complete benchmark model with flavor-diagonal CP violation
to illustrate to what extent new physics to address the muon $g-2$ anomaly is probed by searches for the muon EDM
without being troubled with other CP and flavor observables.
In our CP-violating two-Higgs-doublet model, the muon exclusively couples to one Higgs doublet.
We found that the model has a parameter space which can explain the muon $g-2$ anomaly. 
The parameter region with $m_H = 650$ GeV, $\Delta m_H \sim 320$ GeV, $t_{\beta} \sim 3700$ and $s_{\theta} \sim 0.35$
is particularly interesting because 
it leads to the muon EDM within the sensitivity of the PSI experiment.
The other parameter space still leads to $d_{\mu} = \mathcal{O}(10^{-23}) \, e$ cm
which is covered by the proposed J-PARC experiment.
Our result strongly encourages searches for the muon EDM
which might be a key to open the door of new physics.

\section*{Acknowledgements}

We would like to thank Kim Siang Khaw for inspiring us to work on the muon EDM
and providing useful references.
YN is supported by Natural Science Foundation of China under grant No.~12150610465.

\vspace{0.5cm}

\section*{SUPPLEMENTAL MATERIALS}

{\bf Theoretical conditions.--}
The dimensionless couplings $\lambda_{1,2,3,4},{\rm Re} \lambda_5,{\rm Im} \lambda_5$ in the Higgs potential 
are approximately written as
\begin{align}
\lambda_1 v^2 &\simeq m_h^2 + \left( m_H^2 + s_{\theta}^2 \Delta m_H^2 - M^2 \right) t_{\beta}^2 - 2 \delta_{\tilde h \tilde H} t_{\beta} \, , \label{lambda1} \\[1ex]
\lambda_2 v^2 &\simeq m_h^2 + \left( m_H^2 + s_{\theta}^2 \Delta m_H^2 - M^2 \right) t_{\beta}^{-2} + 2 \delta_{\tilde h \tilde H} t_{\beta}^{-1} \, , \label{lambda2} \\[1ex]
\lambda_3 v^2 &\simeq m_h^2 - m_H^2 - s_{\theta}^2 \Delta m_H^2 - M^2 + 2 m_{H^{\pm}}^2 \nonumber \\
&\hspace{10.0em}+ \delta_{\tilde h \tilde H} \left( t_{\beta}^{-1} - t_{\beta} \right) \label{lambda3} \, , \\[1ex]
\lambda_4 v^2 &\simeq M^2 + m_H^2 + c_{\theta}^2 \Delta m_H^2 - 2 m_{H^{\pm}}^2 \, , \label{lambda4} \\[1ex]
{\rm Re} \lambda_5 v^2 &\simeq M^2 - m_H^2 - c_{\theta}^2 \Delta m_H^2 \, , \label{lambda5} \\[1ex]
{\rm Im} \lambda_5 v^2 &\simeq - \frac{s_{2 \theta}}{c_{\beta}^2 - s_{\beta}^2} \Delta m_H^2 \, . \label{eq:Imlam5tb}
\end{align}
These couplings must satisfy theoretical conditions such as vacuum stability, perturbative unitarity and
the absence of Landau pole at a low scale.

The vacuum stability condition gives
\cite{Deshpande:1977rw,Klimenko:1984qx,Sher:1988mj,Nie:1998yn,Kanemura:1999xf,Arhrib:2010ju}
\begin{align}
&\lambda_1 > 0 \, , \label{eq:vacl1} \\[1ex]
&\lambda_2 > 0 \, , \label{eq:vacl2} \\[1ex]
&\sqrt{\lambda_1 \lambda_2} + \lambda_3 > 0 \, , \label{eq:vacl3} \\[1ex]
&\sqrt{\lambda_1 \lambda_2} + \lambda_3 + \lambda_4 - | \lambda_5 | > 0 \, . \label{eq:vacl4}
\end{align}
The perturbative unitarity condition leads to \cite{Kanemura:1993hm,Akeroyd:2000wc,Ginzburg:2005dt,Arhrib:2010ju,Kanemura:2015ska}
\begin{align}
&\left| \frac{3 (\lambda_1 + \lambda_2) \pm \sqrt{9 (\lambda_1 - \lambda_2)^2 + 4 (2 \lambda_3 + \lambda_4)^2}}{2} \right| < 8 \pi \, , \label{eq:pert1} \\[1ex]
&\left| \frac{(\lambda_1 + \lambda_2) \pm \sqrt{(\lambda_1 - \lambda_2)^2 + 4 \lambda_4^2}}{2} \right| < 8 \pi \, , \label{eq:pert2} \\[1ex]
&\left| \frac{(\lambda_1 + \lambda_2) \pm \sqrt{(\lambda_1 - \lambda_2)^2 + 4 | \lambda_5 |^2}}{2} \right| < 8 \pi \, , \label{eq:pert3} \\[1ex]
&\Bigl| \lambda_3 + 2 \lambda_4 \pm | \lambda_5 | \Bigr| < 8 \pi \, , \label{eq:pert4} \\[1ex]
&\Bigl| \lambda_3 \pm \lambda_4 \Bigr| < 8 \pi \, , \label{eq:pert5} \\[1ex]
&\Bigl| \lambda_3 \pm | \lambda_5 | \Bigr| < 8 \pi \, . \label{eq:pert6}
\end{align}
These conditions constrain the quartic couplings to $\mathcal{O}(1)$ values.

{\bf Loop functions.--}
$I_{S, P, C}(r)$ in Eqs.~\eqref{eq:g-2neutral} and \eqref{eq:g-2charged} are loop functions for $\Delta a_{\mu}$, defined by
\begin{align}
I_S(r) &\equiv \int_0^1 \! dx \frac{x^2 (2 - x)}{r x^2 - x + 1} \, , \label{eq:IS(r)} \\[1ex]
I_P(r) &\equiv \int_0^1 \! dx \frac{- x^3}{r x^2 - x + 1} \, , \label{eq:IP(r)} \\[1ex]
I_C(r) &\equiv \int_0^1 \! dx \frac{- x (1 - x)}{r x + 1 - r} \, . \label{eq:IC(r)}
\end{align}
Here, we ignore a tiny neutrino mass in the loop function $I_C(r)$.

$f_0 (r)$ and $I_{\mu} (r)$ in Eqs.~\eqref{eq:muEDM1loop} and \eqref{eq:muEDMBZ} are loop functions for $d_{\mu}$, defined by
\begin{align}
f_0 (r) &= \int_0^1 \! dx \frac{x^2}{r x^2 - x + 1} \, , \\[1ex]
I_{\mu} (r) &= \int_0^1 \! dx \frac{1 - x (1 - x)}{r - x (1 - x)} \ln \left[ \frac{x (1 - x)}{r} \right] \, .
\end{align}

{\bf Parameter optimization.--}
According to Eqs.~\eqref{eq:lam1}-\eqref{eq:Imlam5}, relevant parameters for determining $\lambda_i$ are $\Delta m_H^2$, $\Delta m_{\pm}^2$, $t_{\beta}$, $s_{\theta}$, $X$ and $\delta_{\tilde h \tilde H}$, and hence, the cutoff scale $\Lambda_{\rm cutoff}$ is controlled by these parameters. 
On the other hand, $\Delta a_{\mu}$ and $d_{\mu}$ mainly depend on $m_H^2$, $\Delta m_H^2$, $\Delta m_{\pm}^2$, $t_{\beta}$ and $s_{\theta}$. 
In fact, we have checked that when we change $X$ and $\delta_{\tilde h \tilde H}$ in the following ranges,
\begin{align}
- \frac{m_h^2}{v^2} < X \lesssim 10 \, , \qquad \left| \delta_{\tilde h \tilde H} \right| \lesssim 10 \times \frac{v^2}{t_{\beta}} \, ,
\label{eq:ranges}
\end{align}
the model predictions for $\Delta a_{\mu}$ and $d_{\mu}$ deviate from those with $X = 0$ and $\delta_{\tilde h \tilde H} = 0$
by less than $0.1\%$. 
Here, the lower limit on $X$ can be understood from Eqs.~\eqref{eq:lam1} and \eqref{eq:vacl1}.
Then, we optimize $X$ and $\delta_{\tilde h \tilde H}$ for a fixed parameter set of
$(\Delta m_H^2, \Delta m_{\pm}^2, t_{\beta}, s_{\theta})$
to maximize $\Lambda_{\rm cutoff}$. 
Note that due to a large $y_{\mu}$ in the large $\tan \beta$ limit,
$\lambda_1$ tends to be negative at high energy scales by the RGE.
Therefore, $X$ is favored to be positive to obtain a higher cutoff scale $\Lambda_{\rm cutoff}$. 
Once $X$ and $\delta_{\tilde h \tilde H}$ are optimized appropriately,
$\Lambda_{\rm cutoff}$ is independent of $s_{\theta}$. 
This is because $s_{\theta}$ dependence in $\lambda_3$ can be absorbed into $\delta_{\tilde h \tilde H}$,
and only the absolute value of $\lambda_5$ is relevant not only to theoretical conditions but also to the RGEs. 
The other $\lambda_i$ do not depend on $s_{\theta}$.

The parameters $\Delta m_H^2, \Delta m_{\pm}^2$ are expected to be $\mathcal{O}(v^2)$,
and we allow them to be up to $(2 v)^2$. 
Actually, when we take more larger values, some of $\lambda_i$ become larger than $\mathcal{O}(1)$,
which results in a lower cutoff scale. 
We choose $\Delta m_{\pm}^2 = \Delta m_H^2 / 2$ which leads to $\lambda_4 \simeq 0$ (see Eq.~\eqref{eq:lam4})
at a low energy scale so that it helps to enhance $\Lambda_{\rm cutoff}$. 
Note that although $\Delta m_{\pm}^2$ is directly related to the charged Higgs mass,
its contribution to $\Delta a_{\mu}$ is sub-dominant. 
We have checked that the prediction for $\Delta a_{\mu}$ is not significantly affected
when $\Delta m_{\pm}^2 \simeq \mathcal{O}(v^2)$. 
We take a large $t_{\beta}$ of $\mathcal{O}(1000)$, 
but it should be smaller than 5000 
to avoid a too large $y_{\mu}$. 
With the above parameter choices,
we calculate the model predictions for $\Delta a_{\mu}$ and $d_{\mu}$
in the range of $0 < s_{\theta} \lesssim 0.7$ and find the maximal value of the cutoff scale $\Lambda_{\rm cutoff}$.

{\bf $T$ parameter.--}
We here check whether the model is consistent with a precision electroweak test, the $T$ parameter
\cite{Peskin:1990zt,Peskin:1991sw}.
A new physics contribution to the $T$ parameter is roughly constrained as $|\Delta T| \lesssim 0.2$
\cite{ParticleDataGroup:2020ssz}. 
The contribution of the 2HDM can be found in refs.~\cite{Grimus:2007if,Grimus:2008nb}.
It mainly depends on mass differences among $H_1$, $H_2$ and $H_{\pm}$. 
In the parameter space of interest,
the mass differences are found to be small enough to satisfy the $T$ parameter constraint. 
For instance, taking $m_H = 650$ GeV, $\Delta m_H = 320$ GeV and $s_{\theta} = 0.35$,
we obtain $\Delta T \simeq -0.03$.

{\bf $h \to \mu^+ \mu^-$ decay.--}
The modification of the SM Higgs coupling to the muon may change the rate of the $h \to \mu^+ \mu^-$ decay~\cite{Ferreira:2020ukv}. 
The current LHC constraint on the branching ratio of $h \to \mu^+ \mu^-$ is
\begin{align}
{\rm BR}(h \to \mu^+ \mu^-) &< 4.7 \times 10^{-4} \quad ({\rm ATLAS}) \, , \\
0.8 \times 10^{-4} < {\rm BR}(h \to \mu^+ \mu^-) &< 4.5 \times 10^{-4} \quad ({\rm CMS}) \, ,
\end{align}
at 95\% C.L.~\cite{ATLAS:2020fzp,CMS:2020xwi}.
Using the SM value of the branching ratio, BR$(h \to \mu^+ \mu^-) = 2.17 \times 10^{-4}$ for $m_h = 125.1$ GeV
\cite{LHCHiggsCrossSectionWorkingGroup:2016ypw},
the modification of the Higgs coupling to the muon $\kappa_{\mu}$
($\kappa_{\mu} = 1$ corresponds to the SM) is constrained as
\begin{align}
\left| \kappa_{\mu} \right| &< 1.47 \quad ({\rm ATLAS}) \, , \label{eq:constATLAS} \\[1ex]
0.61 < \left| \kappa_{\mu} \right| &< 1.44 \quad ({\rm CMS}) \, . \label{eq:constCMS}
\end{align}
In our model, $\kappa_{\mu}$ is obtained from the Lagrangian \eqref{eq:LagYuk} as
\begin{align}
\kappa_{\mu} = R_{11} - R_{21} t_{\beta} \simeq 1 + \frac{\delta_{\tilde h \tilde H}}{M^2} t_{\beta} \, .
\end{align}
From Eqs.~\eqref{eq:M2def} and \eqref{eq:lam3}, we can expect that $M^2 \simeq m_H^2$ and $\delta_{\tilde h \tilde H} t_{\beta} = \mathcal{O}(v^2)$, and hence $\kappa_{\mu} \simeq 1 + \mathcal{O}(0.1)$. 
We have numerically checked that for all the parameter space shown in Figs.~\ref{fig:650GeV} and \ref{fig:700GeV},
the constraint on $\kappa_{\mu}$ in Eqs.~\eqref{eq:constATLAS} and \eqref{eq:constCMS} are satisfied.

\phantomsection

\end{document}